\documentclass[preprint,review,12pt]{elsarticle}

\usepackage{amsmath}
\usepackage{amssymb}
\usepackage{amsfonts}
\usepackage{graphicx}
\usepackage{dcolumn}
\usepackage{bm}
\usepackage{hyperref}
\hypersetup{colorlinks=false}
\biboptions{numbers,square}

\begin{document}

\begin{frontmatter}

\title{Landau Quantization in the Spinning Cosmic String Spacetime}

\author{C. R. Muniz}

\address{Grupo de F\'isica Te\'orica (GFT), Universidade Estadual do Cear\'a, UECE-FECLI, Iguatu-CE, Brazil.}


\author{V. B. Bezerra}

\address{Departamento de F\'{i}sica, Universidade Federal da Para\'{i}ba, Caixa Postal 5008, CEP 58051-970, Jo\~{a}o Pessoa-PB, Brazil}


\author{M. S. Cunha}

\address{Grupo de F\'isica Te\'orica (GFT), Universidade Estadual do Cear\'a-UECE, CEP 60740-000, Fortaleza-CE, Brazil}


\begin{abstract}
We analyze the quantum phenomenon arising from the interaction of a spinless charged particle with a rotating cosmic string, under the action of a static and uniform magnetic field parallel to the string. We calculate the energy levels of the particle in the non-relativistic approach, showing how these energies depend on the parameters involved in the problem. In order to do this, we solve the time independent Schr\"odinger equation in the geometry of the spinning cosmic string, taking into account that the coupling between the rotation of the spacetime and the angular momentum of the particle is very weak, such that makes sense to apply the Schr\"odinger equation in a curved background whose metric has an off diagonal term which involves time and space. It is also assumed that the particle orbits sufficiently far from the boundary of the region of closed timelike curves which exist around this topological defect. 
Finally, we find the Landau levels of the particle in the presence of a spinning cosmic string endowed with internal structure, {\it i.e.}, having finite width and uniformly filled with both material and vacuum energies.
\end{abstract}


\end{frontmatter}

\section{Introduction}
\label{intro}
Spinning cosmic strings are stable one-dimensional topological defects, infinitely long and straight, characterized by an angular parameter $\alpha$ that depends on their linear mass density $\mu$ and by a linear density of angular momentum $\textbf{J}$. These structures are the stationary counterpart of the static cosmic strings, in which $\alpha=1-4G\mu$, probably arising during the very early stages of the known universe, when the first decoupling between the fundamental interactions described by the Standard Model of Elementary Particles occurred \cite{Vilenkin}.

Such objects were studied for the first time as vacuum solutions of General Relativity through the Kerr spacetime reduced to $1+2$ dimensions \cite{Deser}. Shortly after, these three-dimensional solutions were naturally extended to the 4-dimensional spacetime \cite{Gott} in order to describe spinning cosmic strings. Like the static ones, their geometry is locally flat, but not globally. There is a singularity which coincides with the localization of the string, along which the Riemann curvature is infinity.

The spacetime of spinning cosmic strings has peculiar and nontrivial topological properties, which arise from its conical geometry, as well as from the rotation of the spacetime. Like in the case of their static analogues \cite{vilenkin,celio}, these properties result in some interesting phenomena, as gravitomagnetism and the gravitational Aharanov-Bohm effect, both arising at a purely classical level ({\it i.e.}, non-quantum) \cite{Barros}.

Spinning cosmic strings can also present internal structure \cite{Jensen}, and are surrounded by an exotic region that allows the formation of closed timelike curves (CTC's)\cite{Jackiw}, which are problematic from the point of view of the violation of causality. The boundary that defines this region is at a distance proportional to $J/\alpha$ measured from the cosmic string, and provides a natural boundary condition for the involved fields. The spinning cosmic string was also studied in the Cartan-Einstein's theory \cite{Soleng,Ozdemir} and teleparallel gravity \cite{Andrade}, in which the regions of CTC's were also examined. There are also studies on these objects in the context of extra dimensions, including its causal structure and raising criticisms about the possibility of the existence of the region of CTC's \cite{Slagter}.

About the Landau levels of particles localized in the spacetime of spinning cosmic string, there is little literature \cite{Mostafazedeh}, as opposed to what happens with static cosmic strings \cite{Marques,Bakke,Bakke1,eugenio}, in despite of the occurrence of analogues to the spinning cosmic string in condensed matter physics, more specifically vortices in superfluids \cite{shelard,volovik}. We think that the present paper gives a contribution in the direction to suppress this gap, helping to elucidate the theoretical Landau quantization of charged particles placed in the spacetime of spinning cosmic strings, including the possibility of these having internal structure. The approach consists in the resolution of the Schr\"odinger equation in the appropriate metric, considering an approximation in which the particle is at a safe distance
from the boundary of the CTC's region. The levels of energy are extracted from the found solutions. Finally, we will analyze the problem of the Landau levels in the spacetime of a spinning cosmic string with internal structure, whose interior region is uniformly filled with both matter and vacuum energies, the latter represented by a cosmological constant.

This paper is organized as follows: section 2 details the methodology, section 3 presents the results and discuss them and section 4 closes the paper with the conclusions.

\section{Spacetime metric and solution of the Schr\"odinger equation}
\label{sec:1}

The spacetime generated by a spinning cosmic string without internal structure, which is termed ideal spinning cosmic string, is described by the metric given by \cite{clement}
\begin{eqnarray}\label{02}
ds^2=-(cdt+ad\phi)^2+d\rho^2+\alpha^2\rho^2d\phi^2+dz^2 \hspace{1.3cm}\nonumber\\
=-c^2dt^2\!-\!2acdtd\phi\!+\!(\alpha^2\rho^2\!-a^2)d\phi^2+d\rho^2+dz^2,\!
\end{eqnarray}
where the angular deficit parameter that depends on the linear mass density of the cosmic string is $\alpha\leq1$, and $a=4GJ/c^3$ is the rotational parameter of the string, which has unit of distance.

Note that in this case, the source of the gravitational field relative to a spinning cosmic string possesses angular momentum and the metric
(\ref{02}) has an off diagonal term involving time and space. The non-relativistic approach using the Schr\"odinger equation involves only the covariant Laplacian and this implies that such equation can not be applied in a general situation. However, if we consider that the coupling between the angular momentum of the spinning cosmic string and the angular momentum of the particle is very weak, coupling that can be achieved by imposing that $a\Omega=(4GJ/c^3) d\phi/dt \ll 1$, where $\Omega=d\phi/dt$ is the angular velocity of the particle, then we can neglect the time-space component of the metric and use Schr\"odinger equation. On the other hand, if this condition is not satisfied, the Schr\"odinger equation cannot be applied from the beginning.

Thus, taking into account the restrictions imposed on the problem, we will analyze it from the point of view of non-relativistic quantum mechanics in curved spaces by solving the time independent Schr\"odinger equation in a curved background, which is written as
\begin{eqnarray}\label{01}
\widehat{H}\Psi&=&\left[-\frac{\hbar^2}{2m}\frac{1}{\sqrt{g}}\partial_i(\sqrt{g}g^{ij}\partial_j)\right]\Psi=E\Psi,
\end{eqnarray}
where $g^{ij}$ is the inverse of the spatial part of the metric tensor, with $g$ being the determinant of the metric. The term in brackets is proportional the covariant Laplacian operator in arbitrary curved spacetimes. Einstein's summation convention is used here.

The electromagnetic coupling is implemented in the usual way via $-i\hbar\partial_i\rightarrow-i\hbar\partial_i-(e/c)A_i$ (minimal coupling), where $c$ is the speed of light and $e$ is the electric charge of the particle. In this problem, the vector potential $A_i$ has only the azimuthal component, consistent with the presence of a uniform magnetic field $B$ parallel to the spinning cosmic string, which lies along the axis $z$. Then, the non-null covariant component of the potential vector has the form $A_\phi=B\rho^2/2$. We also consider that the particle is at a large distance from the boundary which defines the region of CTC's.
	
In these conditions, the complete equation that describes a particle without spin and with charge $e$ coupled to the electromagnetic potential with azimuthal direction, in the metric given by the equation (\ref{02}) is
\begin{eqnarray}\label{03}
-\frac{\hbar^2}{2m}\left[\partial_{\rho}^2+{\frac{\alpha^2\rho}{(\alpha^2\rho^2-a^2)}}\partial_{\rho}+{\frac{1}{\alpha^2\rho^2-a^2}\partial_{\phi}^2} + \partial_z^2\right]\Psi\,\,\,\, \nonumber\\
+{\,\frac{i\hbar eB\rho^2}{2mc\,(\alpha^2\rho^2-a^2)}\partial_{\phi}\Psi +\frac{e^2B^2\rho^4}{8mc^2\,(\alpha^2\rho^2-a^2)}\Psi}=E\Psi,
\end{eqnarray}
  Now, let us assume that the particle is at a safe distance from the boundary of the region of CTC's, which means that $\rho\gg a/\alpha$. Taking into account the symmetries of the spacetime, the wave function can be written as
\begin{equation}\label{04}
\Psi(\phi,z,\rho)\propto\exp{(i\ell\phi)}\exp{(ikz)}R(\rho),
\end{equation}
 determined by both azimuthal and translational (in $z$ direction) symmetries. Then, the equation (\ref{03}) becomes a ordinary differential equation for the radial part $R(\rho)$, given by
\begin{eqnarray}\label{05}
\frac{d^2R}{d\rho^2}+\frac{1}{\rho}\frac{dR}{d\rho} - \left[\left(\frac{\ell^2}{\alpha^2}-\frac{eB\ell a^2}{\hbar c\alpha^4}\right)\frac{1}{\rho^2}\,\right.\hspace{0.8cm}\hspace{1.6cm}\nonumber\\
\left.+\frac{e^2B^2\rho^2}{4\hbar^2c^2\alpha^2}\right]\!R\!+\!\left(\frac{eB\ell}{\hbar\alpha^2 c}\!+\! \frac{2mE}{\hbar^2}{-\frac{e^2B^2a^2}{4\hbar^2c^2\alpha^4}}\!-\!k^2\right)\!R=0,
\end{eqnarray}
in which we kept terms up to $\mathcal{O}(a^2)$. The solution of equation (\ref{05}), physically compatible with the problem, is given in terms of the confluent hypergeometric function $_1F_1(\beta;\gamma;z)$, as
\begin{eqnarray}\label{06}
R(\rho)={C\exp{\left(-\frac{1}{2}\sqrt{A_2}\rho^2\right)}\rho^{\sqrt{A_1}}\times}\hspace{2.7cm}\nonumber\\
\times{_1}F_1\left(\frac{1}{2}+\frac{\sqrt{A_1}}{2}-\frac{A_3}{4\sqrt{A_2}};1+\sqrt{A_1};\sqrt{A_2}\rho^2\right),
\end{eqnarray}
where $C$ is a normalization constant. $A_1,A_2$ and $A_3$ are the constant coefficients which multiply the terms with $\rho^{-2}R$, $\rho^2R$ and $R$, respectively, in the radial equation (\ref{05}).

\section{Results and Discussion}

In this section, we will analyse the results concerning two kinds of spinning cosmic strings: the one without internal structure and the other with internal structure.
\subsection{Ideal spinning cosmic string} The energy eigenvalues $E_{n\ell}$ of the particle are given from the condition of the regularity of the radial function at the infinity. This occurs when the confluent hypergeometric function is polynomial, {\it i.e.}, if its first entry is a negative integer ($\beta=-n$) \cite{Arfken}. So we have
\begin{eqnarray}
\label{07}
E_{n,\ell}\!=\!\frac{\hbar\omega_c}{2\alpha}\!\left(\!2n\!+\!1\!+\!\sqrt{\frac{\ell^2}{\alpha^2}\!-\!\frac{eB\ell a^2}{\hbar c\alpha^4}}\!-\!\frac{\ell}{\alpha}\right) \!\!+\!\frac{\hbar^2k^2}{2m}{+\frac{e^2B^2a^2}{8m c^2\alpha^4}},\nonumber \!\!\!\!\!\!\!\!\!\!\!\!\!\!\!\!\\
\end{eqnarray}
where $\omega_c=\frac{|e|B}{mc}$ is the cyclotron frequency of the particle. The last term of this equation represents the energy associated with the motion of the particle in the direction parallel to the spinning cosmic string. We remark that the energy eigenvalues are always real, provided that the particle has negative charge ({\it i.e.}, $e\rightarrow -|e|$, when both $\ell$ and $B >0$), in order to describe stable orbits around the string. The same is true if the particle carries positive charge, but now $\ell<0$. Since that the term with the square root is greater than $\ell/\alpha$ in equation (\ref{07}), energy remains positive. We consider the result expressed in this equation consistent, since when one turns off the rotation of cosmic string doing $a\rightarrow0$ in this equation, compatible with the weak coupling discussed in previous section, we exactly obtain the Landau levels found in the spacetime of a static cosmic string \cite{Marques}.

Considering the case in which $eBa^2\ll 2\hbar c\ell\alpha^2$, then we can expand the square root in (\ref{07}) in order to have
\begin{equation}\label{08}
\left(\frac{\ell^2}{\alpha^2}-\frac{eB\ell a^2}{\hbar c\alpha^4}\right)^{1/2}\approx\frac{|\ell|}{\alpha}\left(1-\frac{eBa^2}{2\ell\hbar c\alpha^2}\right),
\end{equation}
and, thus, we can state that the Landau levels of the particle in the geometry of spinning cosmic string are now given by
\begin{equation} \label{09}
E_{n,\ell}=E^{stat}_{n,\ell}+\Delta,
\end{equation}
where $\Delta=\frac{e^2B^2a^2}{4mc^2\alpha^4}\left(\frac{1}{2}-\frac{|\ell|}{\ell}\right)$. These energy levels represent a shift from the Landau levels in the geometry of the static cosmic string. Thus we have an analogous of the quadratic Zeeman effect in the Landau levels induced by the coupling between the magnetic field and the spacetime rotation.

\subsection{Spinning Cosmic String with Internal Structure} We can leave the problem more interesting, when consider the spacetime of a spinning cosmic string endowed with internal structure, {\it i.e.}, having a finite width, uniformly filled with both matter and vacuum energies, the latter represented by a cosmological constant $\Lambda$. The metric for a static cosmic string with these characteristics was found in \cite{Novello}, which is based on the model of Gott-Hiscock cosmic string. For the external region of this cosmic string, the metric obtained in the cited work is
\begin{equation}\label{12}
ds^2=-dt^2+d\rho^2+\bar{\alpha}^2\rho^2d\phi^2+dz^2,
\end{equation}
where the angular parameter $\bar{\alpha}$ is written in terms of the usual static cosmic string one as
\begin{equation}\label{13}
\bar{\alpha}=\alpha\left(\frac{\rho_0+\Lambda}{\rho_0}\right),
\end{equation}
where $\rho_0$ is the internal volumetric density of energy of the cosmic string, that is related to $\mu$ by the relation $\rho_0 S=\mu$, where $S$ is its cross section, which in the idealized cosmic string is null. At this point it is interesting to call attention to the fact that $\rho_0$ is associated with the radius of the boundary which defines de CTC's region through the relation $r_{CTC}=a/\overline{\alpha}=a\rho_0/\alpha(\rho_0+\Lambda)$. Thus, considering $S=\pi r_0^2$, where $r_0$ is the radius of the string, and the fact that $\rho_0S=\mu$, we find that
\begin{equation}
 r_{CTC}=a\mu/(1-4G\mu)(\mu+\Lambda\pi r_0^2),
\end{equation}
 provided that $\alpha=1-4G\mu$. Therefore, the radius of the CTC's region depends on the radius of the cosmic string, as well as on its linear mass density, angular momentum and the cosmological constant.

We now will put this cosmic string to rotate, doing the transformation $dt\rightarrow dt+ad\phi$, as seen in (\ref{02}). The exterior metric is now the same of the spinning cosmic string with the angular parameter redefined. Thus, the Landau levels are obtained by the procedure and in the conditions previously discussed, given by
\begin{eqnarray}\label{14}
E_{n,\ell}=\frac{\hbar\omega_c\rho_0}{2\alpha\left(\rho_0+\Lambda\right)}
\!\left[2n+\!1\!+\!\frac{\sqrt{\ell^2-\frac{eB\ell a^2}{2\hbar c\left[\alpha\left(\frac{\rho_0+\Lambda}{\rho_0}\right)\right]^2}}-\ell}{\alpha\left(\frac{\rho_0+\Lambda}{\rho_0}\right)}\right]\hspace{-0.7cm}\nonumber\\
+{\frac{e^2B^2 a^2}{8mc^2\alpha^4}\left(\frac{\rho_0}{\rho_0+\Lambda}\right)^4}+\frac{\hbar^2k^2}{2m}.\,
\end{eqnarray}
From this equation, we note that the internal structure modifies the spectrum, which now also depends on the vacuum energy of the string. Without this energy, the spinning cosmic string behaves as an ideal one.

 \section{Concluding remarks}

In this work, we studied the Landau levels of a charged particle placed in the spacetime of a spinning cosmic string, which is straight and infinitely long. For this purpose, we have solved the Schr\"odinger equation independent of time, considering both the conical geometry and the spacetime rotation associated with this topological defect. The particle was also coupled to an electromagnetic potential with azimuthal direction, in order to make it interact with a static and spatially homogeneous magnetic field, parallel to the cosmic string.

We made an approach consistent with the requirement of the particle must be located at a safe distance from the boundary that defines the region of closed timelike curves near spinning cosmic string. In addition to this restriction, in order to use the Schr\"odinger equation from the beginning, we imposed that the coupling between both the angular momenta of the spinning cosmic string and of the particle is very weak in such a way that we neglected the off-diagonal terms of the metric given by (\ref{02}).

The obtained eigenvalues for the energy of the particle are such that when the angular momentum of the string is turned off, one recovers the energy levels of the same particle in the geometry of static cosmic string presented in the literature. The problem has been solved without considering any restriction with respect to the intensity of the gravitational field generated by cosmic string (measured by the linear mass density and by its rotation).


We have also calculated the Landau levels of a charged particle without spin in the geometry of a spinning cosmic string endowed with an internal structure, according to the model of Gott-Hiscock. This cosmic string has finite width and is uniformly filled with both matter and vacuum energies, the latter represented by a cosmological constant. It was found that, when the cosmological constant vanishes, we recover the Landau levels for the case of spinning cosmic string with no internal structure, meaning that the latter could only be revealed from the presence of its own vacuum energy density.

It is worth commenting that, in general, the off diagonal term of the metric cannot be neglected, and then, this problem must be analyzed in the relativistic context by using the Klein-Gordon equation and taking into account the appropriate non-relativistic limit. In doing this, the obtained Schr\"odinger equation differs from (\ref{03}). In such situation, it is possible to get purely gravitational effects of induction of the Landau levels through the rotation of spacetime, as analysed in \cite{konno} in the context of the Kerr metric. Otherwise, imposing that the coupling of the angular momentum of the source of the gravitational field (spinning cosmic string) and the one corresponding to the particle is weak, as we did in this paper, we can apply the Schr\"odinger equation {\it ab initio}.

In conclusion, we emphasize that the obtained results are valid only with the restrictions adopted, namely, the particle is located at a safe distance from the boundary that defines the region of closed timelike curves and that the coupling between the angular momentum of the spinning cosmic string and the angular momentum of the particle is very weak, in which situation we can neglect the off diagonal terms of the metric given by (\ref{02}) and apply the Schr\"odinger equation. Nevertheless, the results of this paper can, in principle, be used to identify some effects produced by a spinning cosmic string and contribute to the studies of vortices in superfluids, due to the analogy between them.

\section*{Acknowleddgements}
C. R. Muniz would like to thank Universidade Federal da Para\'iba (UFPB) for the kind welcome and CNPq for the Postdoctoral Fellowship. V. B. Bezerra would like to thank CNPq for partial ﬁnancial support.


\begin{thebibliography}{0}

\bibitem{Vilenkin} A. Vilenkin and E. P. S. Shelard, \textit{Cosmic Strings and Other Topological Defects}, 517. Cambridge University Press, Cambridge (1994).
\bibitem{Deser} S. Deser, R. Jackiw and G.'t Hooft, Ann. Phys. (N.Y.), {\bf152}, 220 (1984).
\bibitem{Gott} J.R.Gott and M.Alpert, Gen. Relativ. Gravitation, {\bf16}, 243 (1984).
\bibitem{vilenkin} A. Vilenkin, Ap. J. {\bf L51}, 282 (1984).
\bibitem{celio} C. R. Muniz and V. B. Bezerra, Annals of Physics, {\bf340}, 1, 87 (2014).
\bibitem{Barros} A. Barros, V. B. Bezerra e C. Romero, Modern Physics Letters A, {\bf18}, 37, 2673 (2003).
\bibitem{clement} G. Clement, Ann. Phys. (NY), {\bf201}, 241 (1990).
\bibitem{Jensen} B. Jensen and H.Soleng, Phys. Rev. D, {\bf45}, 3528 (1992).
\bibitem{Jackiw} Deser, S. and Jackiw, R., Comments. Nucl. Part., {\bf20}, 6, 337 (1992).
\bibitem{Soleng} H. Soleng, Gen. Relativ. Gravitation, {\bf24}, 1, 111 (1992).
\bibitem{Ozdemir} N. Ozdemir, Int.J. Mod. Phys. A, {\bf20}, 2821 (2005).
\bibitem{Andrade} L. C. G. Andrade, arXiv:gr-qc/0102094v1 (2001).
\bibitem{Slagter} R. J. Slagter, Phys. Rev. D, {\bf54}, 4873 (1996).
\bibitem{Mostafazedeh} A. Mostafazedeh, J. Phys. A: Math. Gen., {\bf31}, 7829 (1998).
\bibitem{Marques} G. A. Marques \textit{et al.}, J.Phys.A, {\bf34}, 5945 (2001).
\bibitem{Bakke} K. Bakke \textit{et al.}, Phys. Rev. D, {\bf79}, 024008 (2009).
\bibitem{Bakke1} K. Bakke, Braz. J. Phys., {\bf42}, 437 (2012).
\bibitem{eugenio} E. R. F. Medeiros and E. R. B. de Mello, Eur. Phys. J. C {\bf72}, 2051 (2012).
\bibitem{shelard}  R.L. Davis and E.P.S Shellard, Phys. Rev. Lett. {\bf63}, 2021 (1989).
\bibitem{volovik} G. E. Volovik, JETP Lett., {\bf67}, 881 (1998).
\bibitem{Arfken} G. B. Arfken and H. J. Weber, \textit{Mathematical Methods for Physicists}, 1029. Academic Press, London (1995).
\bibitem{Novello} M. Novello and M. C. M. Silva, Phys.Rev.D, {\bf48}, 10 (1993).
\bibitem{konno} K. Konno and R. Takahashi, Phys. Rev. D, {\bf85}, 061502(R) (2012).

\end{thebibliography}
\end{document}